\documentclass[aps,prl,reprint,superscriptaddress]{revtex4-2}

\usepackage{amsmath}
\usepackage{amssymb}
\usepackage{graphicx}
\usepackage{bm}

\begin{document}

\title{Pressure-Invariant Isotope Effect as Evidence for Electronically Driven Intertwined Order in Pr$_4$Ni$_3$O$_{10}$}

%\title{Constant Oxygen-Isotope Shift of the Spin-Density-Wave Transition under Compression in Pr$_4$Ni$_3$O$_{10}$}

\author{Rustem Khasanov}
 \email{rustem.khasanov@psi.ch}
 \affiliation{PSI Center for Neutron and Muon Sciences CNM, 5232 Villigen PSI, Switzerland}

\author{Thomas J. Hicken}
% \email{thomas.hicken@psi.ch}
 \affiliation{PSI Center for Neutron and Muon Sciences CNM, 5232 Villigen PSI, Switzerland}

\author{Igor Plokhikh}
% \email{igor.plokhikh@psi.ch; i.plokhikh@fz-juelich.de}
 \affiliation{PSI Center for Neutron and Muon Sciences CNM, 5232 Villigen PSI, Switzerland}
 \affiliation{TU Dortmund University, Department of Physics, Dortmund, 44227, Germany}

\author{Ekaterina Pomjakushina}
 \affiliation{PSI Center for Neutron and Muon Sciences CNM, 5232 Villigen PSI, Switzerland}

\author{Hubertus Luetkens}
 \affiliation{PSI Center for Neutron and Muon Sciences CNM, 5232 Villigen PSI, Switzerland}

\author{Zurab Guguchia}
 \affiliation{PSI Center for Neutron and Muon Sciences CNM, 5232 Villigen PSI, Switzerland}

\author{Christof W. Schneider}
 \affiliation{PSI Center for Neutron and Muon Sciences CNM, 5232 Villigen PSI, Switzerland}

\author{Dariusz J. Gawryluk}
 \affiliation{PSI Center for Neutron and Muon Sciences CNM, 5232 Villigen PSI, Switzerland}

\date{\today}

% ============================================================
% ABSTRACT
% ============================================================

\begin{abstract}
We report muon-spin rotation measurements of the pressure dependence of the oxygen-isotope ($^{16}$O/$^{18}$O) effect on the spin-density wave (SDW) transition in the trilayer Ruddlesden-Popper nickelate Pr$_4$Ni$_3$O$_{10}$. At ambient pressure, the SDW transition shows a finite isotope shift, with $^{16}T_{\rm SDW}=158.04(5)$~K and $^{18}T_{\rm SDW}=159.81(6)$~K. Under hydrostatic pressure, $T_{\rm SDW}$ decreases linearly at nearly identical rates for the two isotope compositions, ${\rm d}\,^{16}T_{\rm SDW}/{\rm d}p=-4.93(5)$~K/GPa and ${\rm d}\,^{18}T_{\rm SDW}/{\rm d}p=-4.90(7)$~K/GPa, such that the isotope shift remains essentially unchanged under compression. The absence of pressure enhancement of the isotope effect points to a predominantly electronic origin of the SDW transition and is consistent with recent inelastic x-ray scattering results, suggesting a new regime of intertwined order in trilayer RP nickelates, which is stabilized by strong spin interactions.
\end{abstract}

\maketitle

% ============================================================
% INTRODUCTION
% ============================================================

%\section{Introduction}

\noindent \underline{{\it Introduction.}}
The isotope effect is a fundamental probe of lattice involvement in emergent quantum states. By modifying atomic mass without altering electronic filling or crystal symmetry, isotope substitution directly tests the role of lattice dynamics and electron–phonon coupling in stabilizing ordered phases \cite{Maxwell_PR_1950, Reynolds_PR_1951, Olsen_Cryogenics_1963, Shaw_Physrev_1961, Mathias_PR_1963, Bucher_PhysRevA_1967, Fassnacht-PRL_1966, Budko_PRL_2001, Hinks_Nature_2001, Hein_PhysRev_1963, Geballe_PRL_1961, Finnemore_PRL_1962, Gibson_PhysRev_1966, Gebale_IBM-Res_1962, Fowler_PRL_1967, Stritzker_ZPhys_1972, Batlogg_PRL_1987, Faltens_PRL_1987, Batlogg_PRL2_1987, Bourne_PRL_1987, Franck_PRB_1991, Babushkina_PhysicaC_1991, Zhao_Nature_1997, Zech_Nature_1994, Zech_PhysicaB_1996, Khasanov_YPr123_JPCM_2003, Khasanov_PRB_SSOIE_2003, Tallon_PRL_2005, Keller_MaterialsToday_2008, Khasanov_PRL_LEM-OIE_2004, Mao_PRB_2001, Khasanov_PRB_OIE-general_2006, Khasanov_PRL_YPr123_2008, Rischau_PRR_2022, Schlueter_PhysicaC_2001, Liu_nature_2009, Shirage_PRL_2009, Khasanov_NJP_2010, Khasanov_PRB_IE-FeSe_2010, Zhao_nature_1996, Guguchia_PRL_2014, Medarde_PRL_1998, Luetkens_JMMM_2007, Amit_AdvCondMat_2011, Khasanov_OIE-La327_PRR_2026, Khasanov_Pressure-OIE_La4310_PRR_2026, Shengelaya_PRL_1999, Zhao_PRB_1994, Lanzara_JPCM_1999, Guguchia_PRB_2015, Bendele_PRB_2017}. In correlated-electron materials, isotope studies have proven particularly powerful for disentangling the interplay between charge, spin, and lattice degrees of freedom \cite{Zhao_nature_1996, Guguchia_PRL_2014, Medarde_PRL_1998, Luetkens_JMMM_2007, Amit_AdvCondMat_2011, Khasanov_OIE-La327_PRR_2026, Khasanov_Pressure-OIE_La4310_PRR_2026, Shengelaya_PRL_1999, Zhao_PRB_1994, Lanzara_JPCM_1999, Guguchia_PRB_2015, Bendele_PRB_2017}.

Layered nickelates have recently emerged as a platform for investigating intertwined density-wave order and unconventional superconductivity \cite{Khasanov_OIE-La327_PRR_2026, Khasanov_Pressure-OIE_La4310_PRR_2026, Sun_Nature_2023, Zhang_NatPhys_2024, Wang_Nature_2024, Liu_NatCom_2024, Zhang_JMST_2024, Li_SciBull_2024, Li_ChinPhysLet_2024, Zhou_MatRadExt_2025, Zhou_arxiv_2024, Sakakibara_PRB_2024, Wang_ChinPhysLett_2024, Pei_arxiv_2024, Wang_PRX_2024, Li_arxiv_2025, Puphal_PRL_2024, Zhu_Nature_2024, Zhang_PRX_2025, Khasanov_NatPhys_La327_2025, Huang_Arxiv_2025,  Khasanov_LaPr327_PRR_2025, Khasanov_La327-1313_Arxiv_2025}. In the Ruddlesden–Popper (RP) series $R_{n+1}$Ni$_n$O$_{3n+1}$, the number of NiO$_2$ layers within each structural block controls the thickness of the correlated NiO$_2$ slab and the hierarchy of intra- and interlayer electronic couplings. Of particular interest are the bilayer ($n=2$) compounds $R_3$Ni$_2$O$_7$ and the trilayer ($n=3$) compounds $R_4$Ni$_3$O$_{10}$ (R = La, Pr), both of which exhibit charge-density wave (CDW) and spin-density wave (SDW) instabilities and develop superconductivity under pressure \cite{Sun_Nature_2023, Zhang_NatPhys_2024, Wang_Nature_2024, Liu_NatCom_2024, Zhang_JMST_2024, Li_SciBull_2024, Li_ChinPhysLet_2024, Zhou_MatRadExt_2025, Zhou_arxiv_2024, Sakakibara_PRB_2024, Wang_ChinPhysLett_2024, Pei_arxiv_2024, Wang_PRX_2024, Li_arxiv_2025, Puphal_PRL_2024, Zhu_Nature_2024, Zhang_PRX_2025, Khasanov_NatPhys_La327_2025, Huang_Arxiv_2025}.

Despite these similarities, the nature of charge-spin coupling differs qualitatively between the two RP families. In bilayer $R_3$Ni$_2$O$_7$, CDW and SDW transitions occur at distinct temperatures, indicating relatively weak coupling between the two order parameters \cite{Khasanov_OIE-La327_PRR_2026, Khasanov_NatPhys_La327_2025, Chen_PRL_2024, Chen_PRR_2025, Dan_SciBull_2025, Kakoi_JPSJ_2024, Chen_NatPhys_2024, Ren_CommPhys_2025, Gupta_NatCom_2025, Luo_CinPhysLett_2025, Wang_InorgChem_2024, Wu_PRB_2001, Liu_SciChina_2023, Seo_InorgChem_1996}. By contrast, in trilayer $R_4$Ni$_3$O$_{10}$, CDW and SDW are strongly intertwined and emerge at the same temperature, pointing to a common instability \cite{Khasanov_Pressure-OIE_La4310_PRR_2026, Zhang_NatCommun_2020, Samarakoon_PRX_2023, Norman_PRB_2025, Jia_PRX_2026}. Their responses to pressure further emphasize this distinction: in trilayer systems, both $T_{\rm CDW}$ and $T_{\rm SDW}$ remain equal and decrease simultaneously under compression \cite{Khasanov_Pressure-OIE_La4310_PRR_2026}, whereas in bilayer compounds, pressure enhances the splitting between the CDW and SDW transitions, with $T_{\rm CDW}$ increasing and $T_{\rm SDW}$ decreasing under compression \cite{Khasanov_NatPhys_La327_2025, Khasanov_LaPr327_PRR_2025}.

Oxygen-isotope effect (OIE) studies provide an additional perspective on this hierarchy. In trilayer La$_4$Ni$_3$O$_{10}$, a finite oxygen-isotope shift is observed at the primary transition where $T_{\rm SDW}=T_{\rm CDW}$, while no measurable isotope effect is detected at the lower-temperature spin-reorientation transition below $T_{\rm SDW}$ \cite{Khasanov_Pressure-OIE_La4310_PRR_2026}. In bilayer La$_3$Ni$_2$O$_7$, the CDW transition exhibits a finite isotope shift while the SDW transition shows no detectable isotope response  \cite{Khasanov_OIE-La327_PRR_2026}. These systematic observations demonstrate that isotope sensitivity of the SDW transition is not intrinsic to magnetism itself but emerges only when spin and charge degrees of freedom are strongly intertwined.

Hydrostatic pressure provides a clean tuning parameter to weaken the SDW state in RP nickelates containing trilayer NiO$_2$ sheets without introducing chemical disorder \cite{Khasanov_Pressure-OIE_La4310_PRR_2026, Khasanov_La327-1313_Arxiv_2025}. In cuprates, suppression of antiferromagnetic order by doping is accompanied by a pronounced enhancement of the oxygen-isotope effect, reflecting increased spin-lattice interplay \cite{Khasanov_PRL_YPr123_2008}.
If pressure in trilayer nickelates plays a role analogous to doping in cuprates, one would expect the isotope response of the SDW transition to evolve under compression.

In this work, we combine oxygen-isotope substitution with $\mu$SR measurements under hydrostatic pressure in Pr$_4$Ni$_3$O$_{10}$. By determining $T_{\rm SDW}$ as functions of pressure for ${}^{16}$O- and ${}^{18}$O-substituted samples, we directly test whether the isotope effect evolves as the intertwined density-wave state is weakened.

% ============================================================
% Oxygen-isotope substitution and sample characterization
% ============================================================

%\section{Oxygen-isotope substitution and sample characterization}

\noindent \underline{\it Sample preparation and experimental techniques.}
Details of the sample preparation, room-temperature x-ray diffraction, thermogravimetric analyses, and the experimental methods employed in the present work -- namely muon-spin rotation/relaxation ($\mu$SR) and Raman spectroscopy -- are provided in the Supplemental Material.
A schematic of the oxygen-isotope substitution setup is presented in Refs.~\onlinecite{Khasanov_OIE-La327_PRR_2026,Khasanov_Pressure-OIE_La4310_PRR_2026,Conder_MatScIng_2001,Conder_PhysicaC_2023}. The $^{18}$O enrichment level in the isotope-substituted sample was determined by mass spectrometry to be $87(1)$\%.

% ============================================================
% Raman Experiments.
% ============================================================

%\section{Raman Experiments}

\begin{figure}[t]
\includegraphics[width=0.7\linewidth]{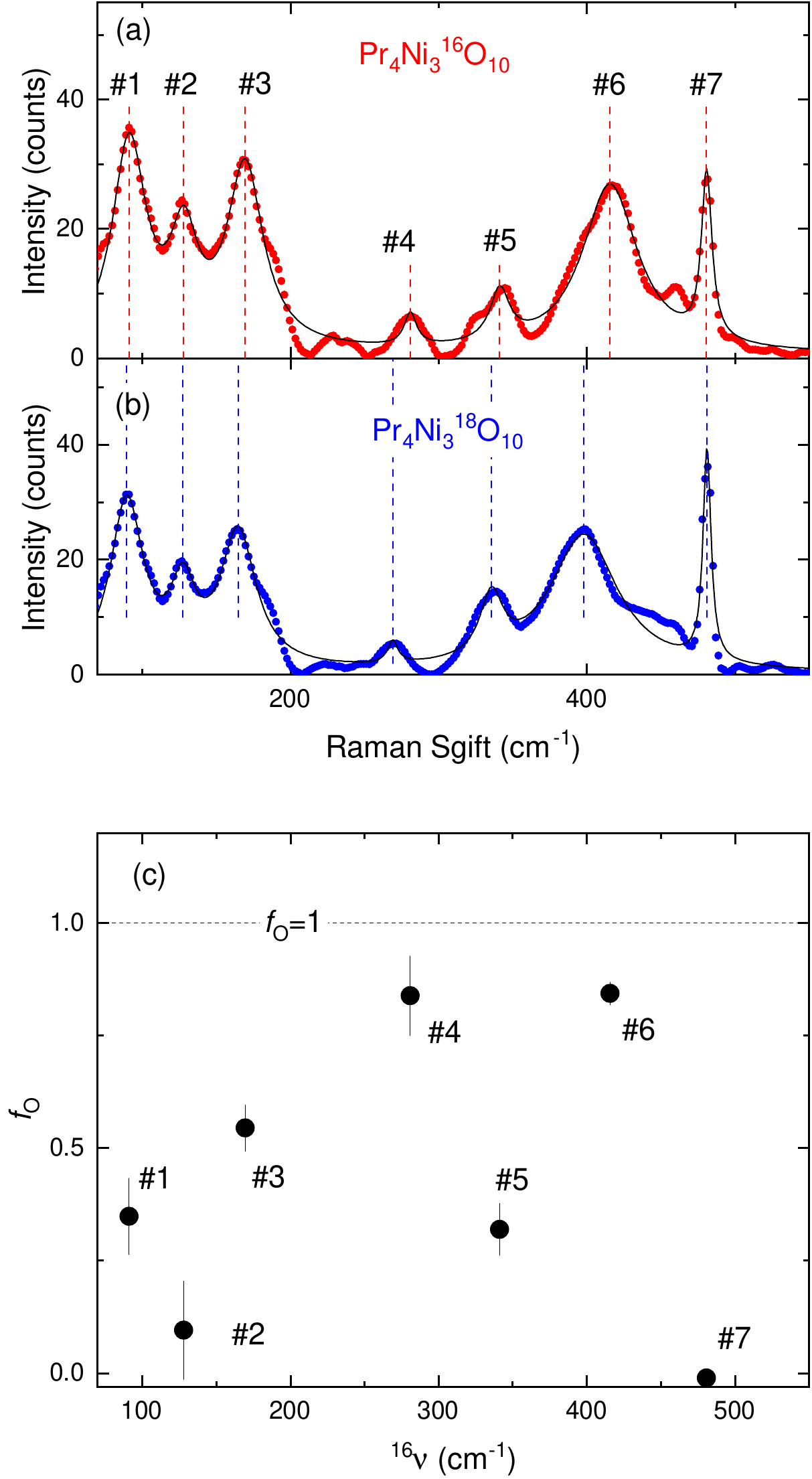}
\caption{
(a) Raman spectra of Pr$_4$Ni$_3$$^{16}$O$_{10}$ (red) and Pr$_4$Ni$_3$$^{18}$O$_{10}$ (blue). Vertical dashed lines indicate the Raman mode positions. Solid lines are Lorenzian fits.
(b) Oxygen participation parameter $f_{\rm O}$ extracted using the partial mass-scaling formalism. The dashed line marks $f_{\rm O}=1$, corresponding to purely oxygen dominated vibrations.
}
\label{fig:Raman}
\end{figure}

\noindent \underline{{\it Raman Experiments.}}
Raman scattering measurements were performed to identify oxygen-related vibrational modes and quantify their oxygen involvement.
Figure~\ref{fig:Raman}(a) displays the Raman spectra of Pr$_4$Ni$_3$$^{16}$O$_{10}$ and Pr$_4$Ni$_3$$^{18}$O$_{10}$ (hereafter the superscripts 16 and 18 denote the $^{18}$O and $^{18}$O substitution, respectively). Seven pronounced Raman-active phonon modes (\#1–\#7) are resolved in the range 80–500~cm$^{-1}$. Upon ${}^{18}$O substitution, most modes exhibit a measurable downshift, consistent with the expected mass renormalization of oxygen-related vibrations.

Because the isotope substitution is partial, the oxygen contribution to each phonon mode was analyzed using the partial mass-scaling formalism developed in Refs.~\onlinecite{Cardona_RMP_2005, Menedez_Philmag_1994, Zhang_PRB_1997}. The isotope dependence of the phonon frequency is described by
\begin{equation}
^{18}\nu(x) = \;^{16}\nu \left[\sqrt{^{16}M/\; ^{18}M(x)}\right]^{f_{\rm O}},
\label{eq:nu_scaling}
\end{equation}
where $x$ is the $^{18}$O fraction ($x=0.87$ in our case), $\nu$ is the phonon frequency, $^{16}M=16$ is the atomic mass of $^{16}$O, $^{18}M(x)=(1-x)\times \; ^{16}M+x\times\; ^{18}M$ is the average oxygen mass in the $^{18}$O substituted sample, and $f_{\rm O}$ is the oxygen participation parameter.

The extracted $f_{\rm O}$ values are shown in Fig.~\ref{fig:Raman}(b). The high-frequency mode \#6 (near 400~cm$^{-1}$) and the intermediate-frequency mode \#4 (near 280~cm$^{-1}$) exhibit $f_{\rm O} \approx 0.9$, consistent with predominantly oxygen character. Modes \#1 to \#3 and mode \#5 display partial oxygen contribution, indicating involvement of heavier ions such as Ni and Pr.
An exception is the highest-frequency mode \#7 near 480~cm$^{-1}$, which does not exhibit any measurable isotope shift within experimental uncertainty. This behavior is unexpected, since high-frequency optical modes in transition-metal oxides generally involve substantial oxygen motion. Moreover, this line displays an anomalously small linewidth compared with other phonon modes, suggesting a different origin. It is therefore likely that this feature is not intrinsic to  Pr$_4$Ni$_3$O$_{10}$ but may arise from a secondary phase, surface contribution, or instrumental artifact.

Apart from this extrinsic feature, no anomalous phonon softening or linewidth broadening is observed. The Raman analysis confirms efficient isotope substitution and identifies the oxygen-dominated vibrational modes. Further theoretical calculations may allow assignment of the observed modes to the corresponding atomic displacements.

\begin{figure*}[t]
\centering
\includegraphics[width=0.8\linewidth]{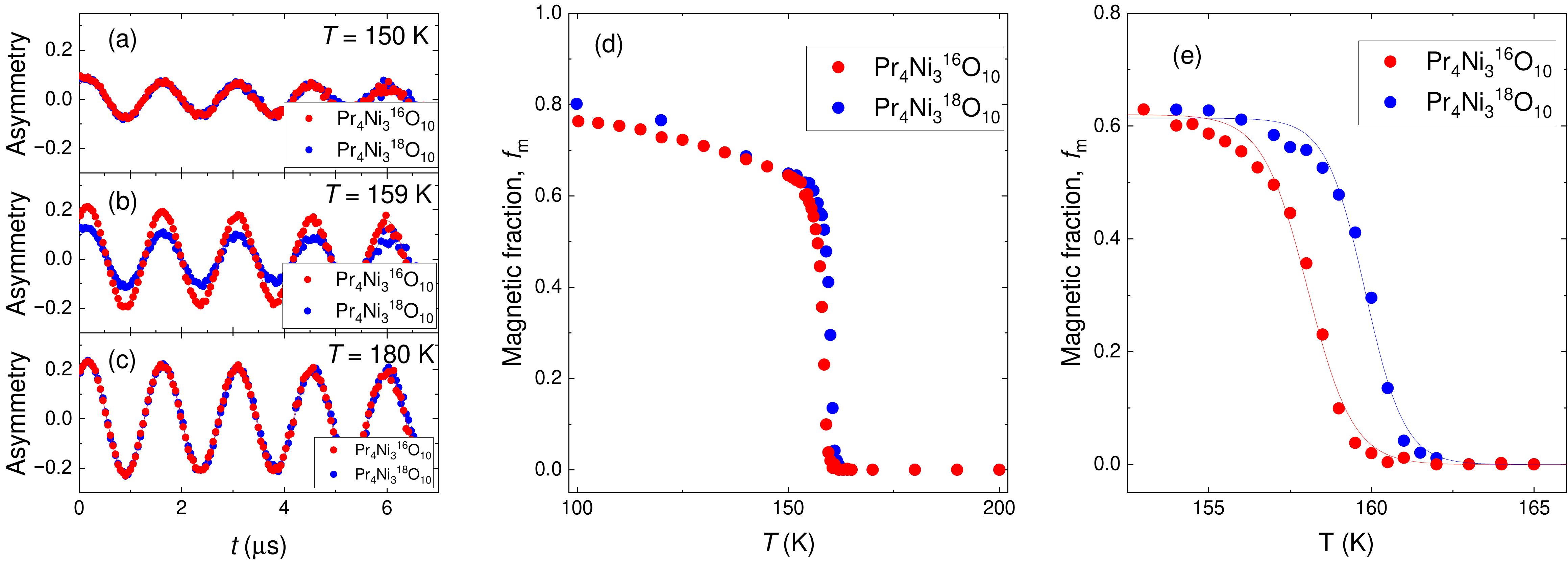}
\caption{
(a)–(c) Weak-transverse-field (WTF) $\mu$SR time spectra measured at ambient pressure in $B_{\rm WTF}=5$~mT at $T=150$~K (a), 159~K (b), and 180~K (c) for the Pr$_4$Ni$_3$$^{16}$O$_{10}$ (red) and Pr$_4$Ni$_3$$^{18}$O$_{10}$  (blue).
(d) Temperature dependence of the magnetic volume fraction $f_{\rm m}$ extracted from the fits.
(e) Enlarged view of the transition region. Solid lines are fits using Eq~\ref{eq:Fermi} with $^{16}T_{\rm SDW}=158.04(5)$~K, $\Delta$$^{16}T_{\rm SDW}=0.69(5)$~K and $^{18}T_{\rm SDW}=159.81(6)$~K, $\Delta$$^{18}T_{\rm SDW}=0.61(6)$~K.
}
\label{fig:WTF_ambient-pressure}
\end{figure*}

% ============================================================
% WTF-muSR: OIE on SDW ordering temperature
% ============================================================

%\subsection{OIE on SDW ordering temperature}

\noindent \underline{{\it OIE on SDW ordering temperature.}}
Recent $\mu$SR and neutron-scattering experiments on Pr$_4$Ni$_3$O$_{10}$ have revealed three magnetic transitions: the first, corresponding to the intertwined CDW and SDW transition at $T_{\rm SDW}\simeq 158$~K; the second, corresponding to a spin-reorientation transition that causes a minor modification of the magnetic structure at $\simeq 90$--100~K; and the third, associated with a major modification of the spin structure due to ordering of the Pr sublattice at $\simeq 25$--27$~$K \cite{Samarakoon_PRX_2023, Khasanov_Pr4310_arxiv_2026}.
In this study, we focus on OIE on the high-temperature SDW transition, which was probed by weak-transverse-field (WTF) $\mu$SR measurements performed in an applied field of $B_{\rm WTF}=5$~mT.

Figure~\ref{fig:WTF_ambient-pressure}~(a)–(c) show representative WTF-$\mu$SR time spectra of Pr$_4$Ni$_3$$^{16}$O$_{10}$ (red symbols) and Pr$_4$Ni$_3$$^{18}$O$_{10}$ (blue symbols) measured at $T=150$~K (a), 159~K (b), and 180~K (c). In the WTF-$\mu$SR configuration, the spins of muons stopping in paramagnetic regions precess coherently in the applied field, whereas muons implanted into magnetically ordered regions experience large static internal fields, leading to rapid depolarization of their spins. The amplitude of the oscillatory component therefore directly reflects the paramagnetic (nonmagnetic) volume fraction of the sample.
Accordingly, at $T=180$~K, well above the SDW transition, both isotope compositions exhibit a full-amplitude oscillatory signal, indicating a fully paramagnetic state. At $T=159$~K, close to the SDW transition, the oscillation amplitude is larger for Pr$_4$Ni$_3$$^{16}$O$_{10}$ than for Pr$_4$Ni$_3$$^{18}$O$_{10}$, implying that the magnetic volume fraction $f_{\rm m}$ is smaller in the ${}^{16}$O-substituted sample than in the ${}^{18}$O-substituted one: ${}^{16}f_{\rm m} < {}^{18}f_{\rm m}$. At $T=150$~K (below the transition), the oscillatory component is strongly suppressed and becomes identical for both isotope compositions, demonstrating that (i) the samples are predominantly in the magnetically ordered state and (ii) the magnetic volume fractions are equal in the two samples: ${}^{16}f_{\rm m} = {}^{18}f_{\rm m}$.

The muon time spectra were analyzed using a model that includes a single oscillatory component representing the response from the paramagnetic (non–magnetically ordered) regions of the sample:
\begin{equation}
A(t)=A_0 (1-f_{\rm m}) \exp\left(-\frac{\sigma_{\rm WTF}^2 t^2}{2}\right)
\cos(\gamma_\mu B_{\rm WTF}\; t+\phi) ,
\label{eq:WTF}
\end{equation}
where $A_0$ is the initial asymmetry of the muon-spin ensemble, $\sigma_{\rm WTF}$ is the Gaussian relaxation rate of the paramagnetic component, $\gamma_\mu = 2\pi \times 135.538$~MHz/T is the muon gyromagnetic ratio, and $\phi$ is the initial phase of the muon-spin ensemble.

The resulting temperature dependence of $f_{\rm m}$ is displayed in Fig.~\ref{fig:WTF_ambient-pressure}~(d). The extension of $f_{\rm m}(T)$ around $T_{\rm SDW}$ is shown Fig.~\ref{fig:WTF_ambient-pressure}~(d). A clear shift of the SDW transition between the two isotope compositions is observed. The solid lines represent fits using a phenomenological Fermi-function form,
\begin{equation}
f_{\rm m}(T)= a
\left[ 1+\exp\!\left(\frac{T-T_{\rm SDW}}{\Delta T_{\rm SDW}}\right) \right]^{-1} +c,
 \label{eq:Fermi}
\end{equation}
where $T_{\rm SDW}$ and $\Delta T_{\rm SDW}$ denote the midpoint and the width of SDW transition and $a$ and $c$ represent the amplitude and the base line offset.

The fits yield $^{16}T_{\rm SDW}=158.04(5)$~K,  $\Delta^{16}T_{\rm SDW}=0.69(5)$~K and $^{18}T_{\rm SDW}=159.81(6)$~K, $\Delta^{18}T_{\rm SDW}=0.61(6)$~K for the Pr$_4$Ni$_3$$^{16}$O$_{10}$ and Pr$_4$Ni$_3$$^{18}$O$_{10}$, respectively. The corresponding isotope shift is $^{18}T_{\rm SDW}-^{16}T_{\rm SDW}=1.77(8)$~K. Within experimental uncertainty, the transition widths are comparable for both isotope compositions, indicating that OIE on $T_{\rm SDW}$ in Pr$_4$Ni$_3$O$_{10}$ is {\it intrinsic}.

\begin{figure*}[t]
\centering
\includegraphics[width=0.9\linewidth]{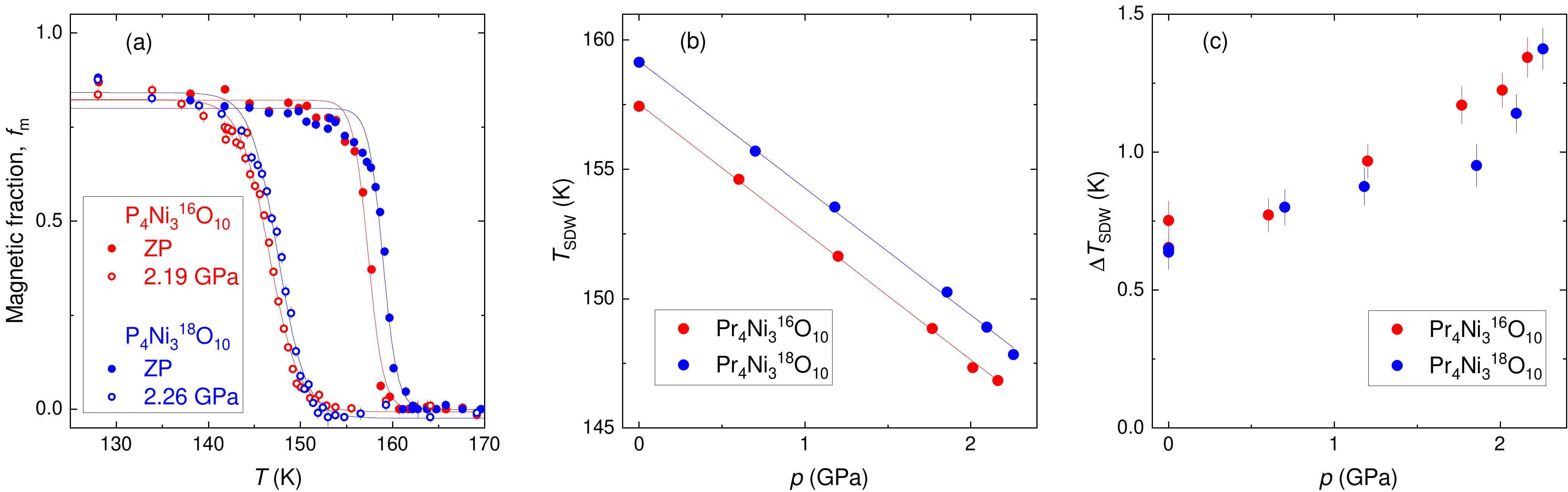}
\caption{
(a) Temperature dependence of the magnetic volume fraction $f_{\rm m}$ for Pr$_4$Ni$_3$$^{16}$O$_{10}$ and Pr$_4$Ni$_3$$^{18}$O$_{10}$ measured at zero pressure (solid symbols) and at the highest applied pressures (open symbols). Solid lines are fits using Eq.~\ref{eq:Fermi}.
(b) Pressure dependence of the SDW transition temperature. Linear fits yield $^{16}T_{\rm SDW}(p)  =  157.52(11)\,{\rm ~K} - p \cdot 4.93(5)\, {\rm ~K/GPa }$ and $^{18}T_{\rm SDW}(p)  =  159.17(10)\,{\rm ~K} - p \cdot 4.90(7)\, {\rm ~K/GPa }$. Within experimental uncertainty, the pressure slopes are identical.
(c) Pressure dependence of the transition width $\Delta T_{\rm SDW}$.
}
\label{fig:WTF_pressure}
\end{figure*}

% ============================================================
% Pressure dependence of OIE on $T_{\rm SDW}$
% ============================================================

%\subsection{Pressure dependence of OIE on $T_{\rm SDW}$}

\noindent \underline{{\it Pressure dependence of OIE on $T_{\rm SDW}$.}}
To investigate the pressure dependence of the oxygen-isotope effect on the SDW transition, WTF-$\mu$SR measurements were performed under hydrostatic pressure for both ${}^{16}$O- and ${}^{18}$O-substituted Pr$_4$Ni$_3$O$_{10}$ samples. In $\mu$SR experiments under pressure, a substantial fraction of the muons ($\simeq 50$\% in our case) are stopped in the pressure cell walls, leading to a significant background contribution \cite{Khasanov_HPR_2016, Shermadini_HPR_2017, Khasanov_JAP_2022, Khasanov_HPR_2022}.

In the analysis of the WTF-$\mu$SR data using Eq.~\ref{eq:WTF}, the pressure cell contribution is included as an additional oscillatory term with a temperature-independent initial asymmetry.
Consequently, the original term $A_0(1-f_{\rm m})$ in Eq.~\ref{eq:WTF} is replaced by $A_{0,{\rm s}} (1-f_{\rm m}) + A_{0,{\rm bg}}$, where the subscripts s and bg denote the sample and background (pressure cell) contributions, respectively. In this case, $\sigma_{\rm WTF}$ accounts for the Gaussian depolarization arising from the paramagnetic response of both the sample and the pressure cell.

Figure~\ref{fig:WTF_pressure}~(a) shows the temperature dependence of the magnetic volume fraction for both isotope compositions at zero pressure and at the highest pressures reached in the experiments (2.19~GPa for the ${}^{16}$O sample and 2.26~GPa for the ${}^{18}$O sample). With increasing pressure, the SDW transition shifts to lower temperatures for both isotopes. The overall shape of the $f_{\rm m}(T)$ curves remains nearly unchanged between the lowest and highest pressures, although the transition width appears to increase slightly under compression.

The pressure dependencies of the SDW transition temperature $T_{\rm SDW}(p)$, extracted from the fit of $f_{\rm m}(T)$ using Eq.~\ref{eq:Fermi}, are plotted in Fig.~\ref{fig:WTF_pressure}~(b). The linear fits reveal:
\begin{equation}
^{16}T_{\rm SDW}(p)  =  157.52(11)\,{\rm ~K} - p \cdot 4.93(5)\, {\rm ~K/GPa } \nonumber
\end{equation}
and
\begin{equation}
^{18}T_{\rm SDW}(p)  =  159.17(10)\,{\rm ~K} - p \cdot 4.90(7)\, {\rm ~K/GPa }.  \nonumber
\end{equation}
The nearly equal slopes  demonstrate that the oxygen-isotope shift of the SDW transition remains {\it constant} under compression.

Figure~\ref{fig:WTF_pressure}~(c) displays the pressure dependence of the transition width $\Delta T_{\rm SDW}$. $\Delta T_{\rm SDW}$ increases nearly equally with pressure for both isotope compositions. This broadening of the transition may be intrinsic; however, it could also be related to increasing nonhydrostatic conditions inside the pressure cell at elevated pressures, leading to a distribution of local pressures and, consequently, to a spread of transition temperatures across the sample. At present, it is not possible to distinguish between these two scenarios.
Importantly, although the transition becomes broader with increasing pressure, the linear suppression of $T_{\rm SDW}$ proceeds with the same rate for both ${}^{16}$O- and ${}^{18}$O-substituted samples.

%Our results imply, therefore, that pressure weakens the SDW state in Pr$_4$Ni$_3$O$_{10}$ but does not modify the magnitude of the oxygen-isotope shift, establishing the pressure independence of the isotope effect on the SDW transition.

% ============================================================
% DISCUSSION
% ============================================================

%\section{Discussion}

\noindent \underline{{\it Discussion and Conclusions.}}
The present results establish two central observations:
\begin{itemize}
    \item[1.] A finite oxygen-isotope shift of the SDW transition temperature, with $^{18}T_{\rm SDW}-^{16}T_{\rm SDW}=1.77(8)$~K; and
    \item[2.] An identical pressure dependence of $T_{\rm SDW}$ for both isotope compositions, with ${\rm d}T_{\rm SDW}/{\rm d}p = -4.92(7)$~K/GPa.
\end{itemize}
These findings imply that although pressure suppresses the SDW phase, the magnitude of the OIE on $T_{\rm SDW}$ in Pr$_4$Ni$_3$O$_{10}$ remains independent of pressure.

Following previous isotope studies on bilayer La$_3$Ni$_2$O$_7$
\cite{Khasanov_OIE-La327_PRR_2026} and trilayer La$_4$Ni$_3$O$_{10}$
\cite{Khasanov_Pressure-OIE_La4310_PRR_2026},
the isotope sensitivity of the SDW order appears only when spin and charge degrees of freedom are strongly intertwined.
The pressure dependence provides a decisive test of this picture.
If lattice involvement were to increase as the intertwined CDW/SDW instability weakens -- particularly as the system approaches the superconducting regime -- one would expect the isotope response of $T_{\rm SDW}$ to evolve.
Instead, the isotope shift remains unchanged over the entire investigated pressure range.
This demonstrates that suppression of the intertwined CDW/SDW state under compression does not enhance lattice participation in the density-wave ordering mechanism.

This conclusion is consistent with recent inelastic x-ray scattering results by Jia {\it et al.}~\cite{Jia_PRX_2026},
who investigated lattice--charge coupling in trilayer RP nickelates La$_4$Ni$_3$O$_{10}$ and Pr$_4$Ni$_3$O$_{10}$.
In contrast to canonical cuprate superconductors -- where the onset of intertwined CDW/SDW order is accompanied by pronounced phonon softening near the ordering wave vector
\cite{McQueeney_PRL_1999, Reznik_Nature_2006, Baron_JPCS_2008, Le_Tacon_NatPhys_2014, Miao_PRX_2018} -- no measurable phonon softening was detected across the density-wave transition in the trilayer nickelates.
Moreover, calculations of the electronic susceptibility revealed a pronounced peak at the SDW wave vector but not at the CDW wave vector, emphasizing the dominant role of spin correlations.
Together with our observation of a pressure-independent isotope shift, these results indicate that the density-wave instability in trilayer nickelates is primarily electronic in origin, with lattice degrees of freedom playing a secondary, cooperative role.

The behavior contrasts sharply with that of cuprate superconductors.
In cuprates, suppression of antiferromagnetic order by doping is accompanied by a substantial enhancement of the oxygen-isotope effect on superconductivity, reflecting strong spin–lattice interplay \cite{Khasanov_PRL_YPr123_2008, Guguchia_PRL_2014, Shengelaya_PRL_1999, Zhao_PRB_1994, Lanzara_JPCM_1999, Guguchia_PRB_2015, Bendele_PRB_2017}.
If pressure in trilayer nickelates plays an analogous role -- weakening magnetic order and moving the system toward superconductivity -- one might anticipate an increasing lattice contribution to the relevant energy scale.
The absence of such enhancement in Pr$_4$Ni$_3$O$_{10}$ suggests a fundamental difference between the two material classes.

To conclude, the finite but pressure-invariant oxygen-isotope shift, combined with the absence of phonon anomalies reported by Jia {\it et al.} \cite{Jia_PRX_2026}, supports a scenario in which electronic correlations dominate the intertwined CDW/SDW transition in trilayer RP nickelates.
While the lattice participates in the ordering process, as evidenced by the finite isotope effect, it does not control the pressure evolution of the SDW state.
These findings place important constraints on microscopic models of density-wave formation and on proposed mechanisms of superconductivity in Ruddlesden–Popper nickelates.

\end{document}